\documentclass[12pt]{article}
\usepackage{graphicx}

\def\la{\mathrel{\mathchoice {\vcenter{\offinterlineskip\halign{\hfil%
$\displaystyle##$\hfil\cr<\cr\sim\cr}}}{\vcenter{\offinterlineskip%
\halign{\hfil$\textstyle##$\hfil\cr<\cr\sim\cr}}}{\vcenter{%
\offinterlineskip\halign{\hfil$\scriptstyle##$\hfil\cr<\cr\sim\cr}}}%
{\vcenter{\offinterlineskip\halign{\hfil$\scriptscriptstyle##$\hfil\cr%
<\cr\sim\cr}}}}}
\def\ga{\mathrel{\mathchoice {\vcenter{\offinterlineskip\halign{\hfil%
$\displaystyle##$\hfil\cr>\cr\sim\cr}}}{\vcenter{\offinterlineskip%
\halign{\hfil$\textstyle##$\hfil\cr>\cr\sim\cr}}}{\vcenter{%
\offinterlineskip\halign{\hfil$\scriptstyle##$\hfil\cr>\cr\sim\cr}}}%
{\vcenter{\offinterlineskip\halign{\hfil$\scriptscriptstyle##$\hfil\cr%
>\cr\sim\cr}}}}}

\begin{document}
\def\baselinestretch{1.5}\large\normalsize

\begin{flushright}
Accepted for publication in Icarus, November 1, 2007
\end{flushright}

\begin{center}
{\bf {\large Fragment-Collision Model for Compound Chondrule Formation: Estimation of Collision Probability}}

\vspace{2mm}
{Hitoshi Miura$^{1,2,3}$, Seiji Yasuda$^{2,4,5}$, and Taishi Nakamoto$^{4}$}

\vspace{10mm}
\small
\noindent $^1${\it Theoretical Astrophysics Group, Department of Physics, Kyoto University, Kitashirakawa, Sakyo, Kyoto 606-8502, Japan}

\noindent $^2${\it Research Fellow of the Japan Society for the Promotion of Science}

\noindent $^3${\it Corresponding Author E-mail address: miurah@tap.scphys.kyoto-u.ac.jp}

\noindent $^4${\it Earth and Planetary Sciences, Tokyo Institute of Technology, Meguro, Tokyo 152-8551, Japan}

\noindent $^5${\it Pure and Applied Sciences, University of Tsukuba, 1-1-1 Tenno-dai, Tsukuba 305-8577, Japan}
\end{center}

\vspace{2cm}

\noindent Pages: 34

\noindent Tables: 0

\noindent Figures: 8

\def\baselinestretch{1.0}\large\normalsize

\normalsize

\clearpage

\noindent {\bf Proposed Running Head:} New Model for Compound Chondrule Formation

\vspace{2cm}

\noindent {\bf Editorial correspondence to:}

\noindent Dr. Hitoshi Miura

\noindent Theoretical Astrophysics Group, Department of Physics, Kyoto University

\noindent Kitashirakawa, Sakyo, Kyoto 606-8502, Japan

\noindent Phone: +81-75-753-3885

\noindent Fax: +81-75-753-3886

\noindent E-mail: miurah@tap.scphys.kyoto-u.ac.jp

\clearpage

\abstract{
We propose a new scenario for compound chondrule formation named as ``fragment-collision model," in the framework of the shock-wave heating model. A molten cm-sized dust particle (parent) is disrupted in the high-velocity gas flow. The extracted fragments (ejectors) are scattered behind the parent and the mutual collisions between them will occur. We modeled the disruption event by analytic considerations in order to estimate the probability of the mutual collisions assuming that all ejectors have the same radius. In the typical case, the molten thin ($\sim 1 \, {\rm mm}$) layer of the parent surface will be stripped by the gas flow. The stripped layer is divided into about 200 molten ejectors (assuming that the radius of ejectors is $300 \, {\rm \mu m}$) and then they are blown away by the gas flow in a short period of time ($\sim 0.01 \, {\rm s}$). The stripped layer is leaving from the parent with the velocity of $\sim 4 \, {\rm cm \, s^{-1}}$ depending on the viscosity, and we assumed that the extracted ejectors have a random velocity $\Delta v$ of the same order of magnitude. Using above values, we can estimate the number density of ejectors behind the parent as $n_{\rm e} \sim 800 \, {\rm cm^{-3}}$. These ejectors occupy $\sim 9\%$ of the space behind the parent in volume. Considering that the collision rate (number of collisions per unit time experienced by an ejector) is given by $R_{\rm coll} = \sigma_{\rm coll} n_{\rm e} \Delta v$, where $\sigma_{\rm coll}$ is the cross-section of collision (e.g., Gooding \& Keil 1981, Meteoritics 16, 17), we obtain $R_{\rm coll} \sim 36 \, {\rm collisions/s}$ by substituting above values. Since most collisions occur within the short duration ($\sim 0.01 \, {\rm s}$) before the ejectors are blown away, we obtain the collision probability of $P_{\rm coll} \sim 0.36$, which is the probability of collisions experienced by an ejector in one disruption event. The estimated collision probability is about one order of magnitude larger than the observed fraction of compound chondrules. In addition, the model predictions are qualitatively consistent with other observational data (oxygen isotopic composition, textural types, and size ratios of constituents). Based on these results, we concluded that this new model can be one of the strongest candidates for the compound chondrule formation. 

It should be noted that all collisions do not necessarily lead to the compound chondrule formation. The formation efficiency and the future works which should be investigated in the forthcoming paper are also discussed. 
}\\


\clearpage

\section{Introduction}
\label{sec:introduction}
Chondrules are millimeter-sized, once-molten, spherical-shaped grains mainly composed of silicate material. They are abundant in chondritic meteorites, which are the majority of meteorites falling onto the Earth. They are considered to have formed from chondrule precursor dust particles about $4.56\times10^9 \, {\rm yr}$ ago in the solar nebula (Amelin et al. 2002); they were heated and melted through flash heating events in the solar nebula and cooled again to solidify in a short period of time (e.g., Jones et al. 2000, and references therein). Typical chondrules are single spherical objects, while compound chondrules are composed of two or more chondrules fused together. They are rare in all chondrules ($\sim 4 \%$), but occur in many classes of chondrites, so they offer crucial information regarding the physical and chemical state of solid materials during chondrule formation (Gooding \& Keil 1981, Wasson et al. 1995, Sekiya \& Nakamura 1996, Ciesla et al. 2004, Akaki \& Nakamura 2005). 

Wasson et al. (1995) defined compound chondrules based on textures and assigned each constituent chondrule as primary or secondary. The primary chondrule was rigid enough to retain its original shape at the time of compound chondrule formation. On the contrary, the secondary chondrule had a low viscosity enough to allow it to conform the shape of the primary. Akaki \& Nakamura (2005) considered four types of compound chondrules as follows: (1) the enveloping type, wherein a secondary chondrule encloses a primary one, (2) the adhering type, wherein a small secondary chondrule forms a hemispherical bump on the surface of a larger primary chondrule, (3) the consorting type, wherein the conjoined chondrules are of similar size, and (4) the blurred type, wherein compound chondrules have a blurred boundary between two chondrules, which makes it difficult to define the primary or the secondary. They classified 30 compound chondrule sets found in two CV3 chondrites into 7 blurred, 13 adhering, 4 consorting, and 6 enveloping. 

At least four models have been proposed for compound chondrule formation. (1) Random collision model; totally or partially molten particles randomly collided in the solar nebula (Gooding \& Keil 1981, Sekiya and Nakamura 1996). (2) Collision between individual chondrules in shock-wave heating (Ciesla 2006). (3) Eruption model; immediately after the formation of the solidified shell in the primary chondrule, the inner residual melt was vacuumed out through a crack of the shell, and formed the secondary chondrule on the surface of the primary (Sanders \& Hill 1994). (4) Relict grain model; fine-grained dust particles accreted on the surface of already-formed primary were melted in second heating event (Wasson 1993, Wasson et al. 1995). The relict grain model is good for enveloping type if considering that fine-grains accreted uniformly on the primary surface, but it seems to be difficult to explain other types. In the eruption model, the secondary was formed from the inner residual melt after the formation of the solidified shell on the primary surface, however, the chemical compositions of some sets of compounds are inconsistent with this scenario (Akaki \& Nakamura 2005). The random collision model is considered not to account for the observed fraction of compound chondrules because of the low density of matter in the nebula (Gooding \& Keil 1981, Wasson et al. 1995, Sekiya and Nakamura 1996). Collision between individual chondrules in shock-wave heating can account for the observed fraction of compound chondrules if the enhancement of the dust particle in the pre-shock region is $\sim 400$ times that expected under canonical conditions (Ciesla 2006). However, such a highly dust-rich region is unfavorable to explain the scarcity of isotopic fractionation of sulfur if the shock wave has a large spatial extent in solar nebula (Tachibana \& Huss 2005, Miura \& Nakamoto 2006). 

In this paper, we propose a new scenario for compound chondrule formation. The shock-wave heating model is one of the most plausible models for chondrule formation (e.g., Connolly \& Love 1998). In this model, the dust particles are exposed to a high-velocity gas flow and heated by the gas frictional heating. It has been suggested that the maximum sizes of chondrules are regulated by the gas flow because large dust particles should be disrupted by the strong gas ram pressure when they melt (Susa \& Nakamoto 2002, Kato et al. 2006). Recently, we carried out three-dimensional hydrodynamics simulations of molten dust particle exposed to the gas flow and showed that molten cm-sized dust particle is disrupted into many small pieces in a typical setting of nebula shocks (Miura \& Nakamoto 2007). These pieces have many chances of mutual collisions to form compound chondrules because the local number density of them behind the disrupted particle is enhanced. We name this scenario ``fragment-collision model" and think that it can be a strong candidate for compound chondrule formation model. This model seems very similar to the model of collision between individual chondrules in the shock-wave heating at the point that compound chondrules are formed in the gas flow (Ciesla 2006). The difference is that in our model, compound chondrules are formed from a single large dust particle, so the dust enhancement in the pre-shock region is not necessarily required. 

The purpose of this paper is to estimate the collision probability, i.e., the number of collisions which a single piece will experience in a disruption event, by using a simple formulation. We call the disrupted dust particle as ``parent" and small pieces as ``ejectors." For simplicity, we assume that all ejectors have the same radius of $r_{\rm e}$ in this paper. We describe the formulations for estimating the collision probability in \S \ref{sec:formulation}. The expected probability is discussed in \S \ref{sec:predicted_collision_frequency}. We consider appropriate situations in which our model can be applied in \S \ref{sec:parameter_range}. We compare our model with observational data of compound chondrules in \S \ref{sec:observation}. Finally, we make conclusions in \S \ref{sec:summary}.

\section{Formulation}
\label{sec:formulation}

\subsection{Collision Rate}
\label{sec:collision_frequency}
\begin{figure}[p]
\center
\includegraphics[scale=.6, angle=0.]{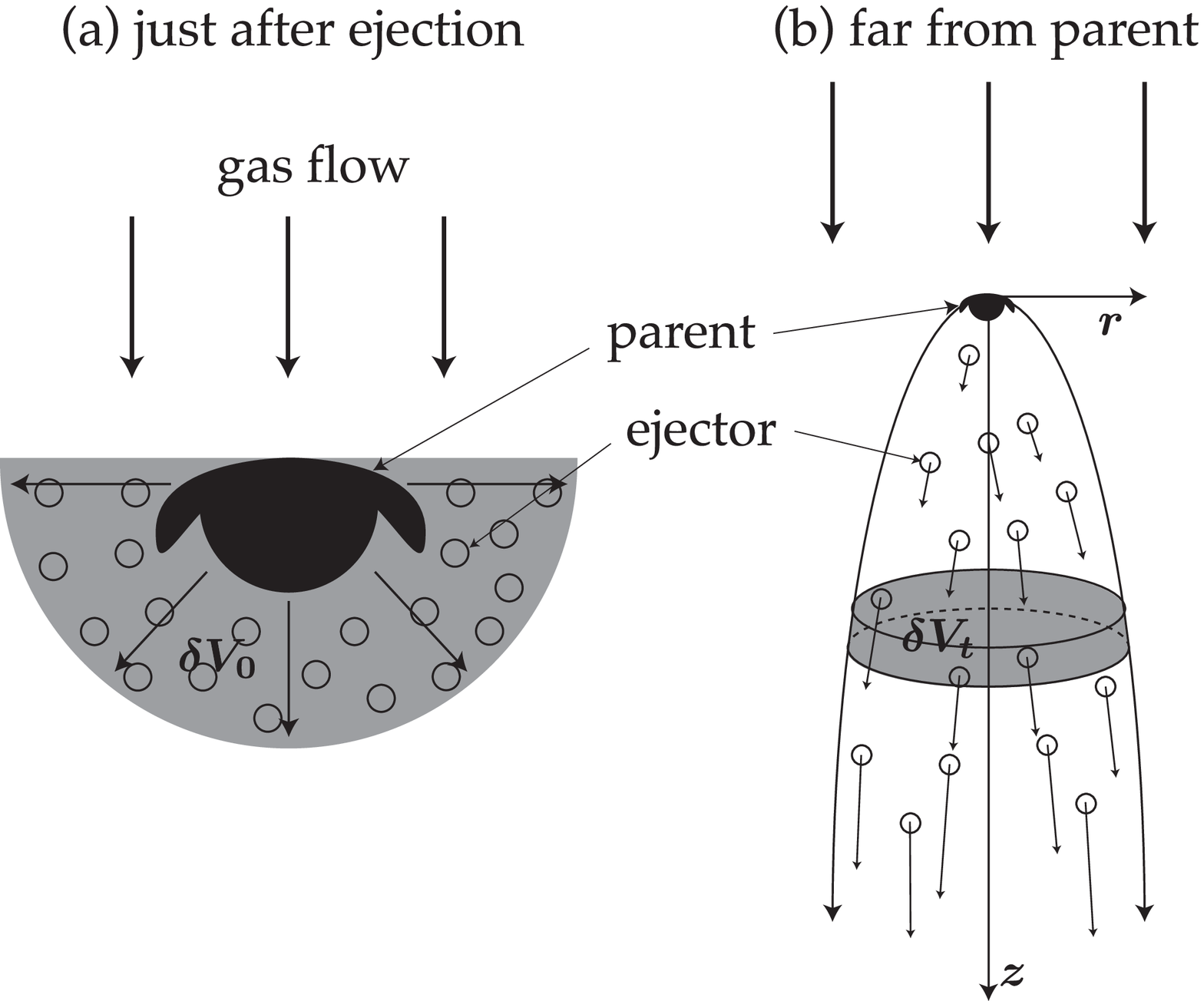}
\caption{Schematic picture of fragmentation of molten chondrule precursor dust particle (parent) in the shock-wave heating event. When the gas ram pressure is too strong for the molten particle to keep its shape, many small particles (ejectors) are dispersed behind the parent particle and mutual collisions between them will occur. (a) Just after ejection, these ejectors part from the surface of the parent at rear side with initial ejection velocity. (b) Far from the parent particle, the motions of ejectors become an uniformly-accelerated motion along the direction of the gas flow ($z$-axis), and a constant-velocity motion perpendicular to the gas flow ($r$-axis). }
\label{fig:volume}
\end{figure}

The number of collisions per unit time experienced by each ejector (collision rate) is given by $R_{\rm coll} = \sigma_{\rm coll} n_{\rm e} \Delta v$, where $\sigma_{\rm coll}$ is the collisional cross-section ($\sigma_{\rm coll} = 4 \pi r_{\rm e}^2$), $n_{\rm e}$ is the number density, and $\Delta v$ is the velocity dispersion of ejectors (Gooding \& Keil 1981, Sekiya \& Nakamura 1996). The original point of our model is to estimate $n_{\rm e}$ resulting from disruption of the parent. Considering the total number of ejectors torn away from the parent in an infinitesimal duration $\delta t$, $\delta N$, and the volume of the region in which these ejectors are scattered, $\delta V$, we obtain $n_{\rm e} = \delta N / \delta V$. Assuming that all ejectors just after ejection are parting from the parent with a velocity of $\sim \Delta v$, we obtain the volume in this phase as $\delta V_0 \sim 2 \pi r_{\rm p}^2 \Delta v \delta t$, where $r_{\rm p}$ is the radius of parent (see Fig. \ref{fig:volume}a). It should be noted that ejectors are jumping out of rear side of the parent surface, not of front one. After ejection, the motions of ejectors are affected by the ambient gas flow. We simply assume that ejectors are accelerated with a constant acceleration $a$ in the direction of the gas flow ($z$-axis), on the other hand, in the direction perpendicular to the gas flow ($r$-axis) they move with a constant velocity of $\sim \Delta v$ (see Fig. \ref{fig:volume}b). The acceleration is given by $a = 3 p_{\rm fm} / 4 r_{\rm e} \rho_{\rm mat}$, where $p_{\rm fm}$ is the gas ram pressure and $\rho_{\rm mat}$ is the material density inside of the molten dust particle. In this later phase, the region in which ejectors are scattered is getting wider steeply with time $t$ and its volume is given by $\delta V_t \sim \pi (\Delta v t)^2 a t \delta t$. Approximating $\delta V \simeq \delta V_0 + \delta V_t$, we obtain the number density of ejectors as
\begin{equation}
n_{\rm e} \sim \left( R_{\rm eject} /  2 \pi r_{\rm p}^2 \Delta v \right) \, \left[ 1 + \left( t / t_{*} \right)^3 \right]^{-1} 
\label{eq:number_density}
\end{equation}
and the collision rate
\begin{equation}
R_{\rm coll} \sim \left( 2 r_{\rm e}^2 R_{\rm eject} / r_{\rm p}^2 \right) \, \left[ 1 + \left( t / t_{*} \right)^3 \right]^{-1} , 
\label{eq:collision_rate}
\end{equation}
where $R_{\rm eject}$ is the total number of ejectors extracted from the parent per unit time (ejection rate) defined by $R_{\rm eject} \equiv \delta N / \delta t$. The time $t_{*}$ is defined as $t_{*} \equiv \left( 8 \rho_{\rm mat} r_{\rm p}^2 r_{\rm e} / 3 \Delta v p_{\rm fm} \right)^{1/3}$ and it gives the timescale within which most collisions will occur (see \S \ref{sec:collision_probability}). 

\hspace{1cm} {\bf [Figure 1]}

\subsection{Collision Probability}
\label{sec:collision_probability}
%The number of collisions experienced by each ejector per a disruption event (collision probability), $P_{\rm coll}$, is given by integrating the collision rate $R_{\rm coll}$ over the time $t$ from $0$ to $\infty$ (see Eq. \ref{eq:collision_rate}). 
The probability of collisions that each ejector experiences during the time from $t_1$ to $t_2$ is given by integrating the collision rate $R_{\rm coll}$ over the time $t$ from $t_1$ to $t_2$ as (see Eq. \ref{eq:collision_rate})
\begin{equation}
P_{\rm coll} (t_1 , t_2) \equiv \int_{t_1}^{t_2} R_{\rm coll} dt .
\end{equation}
In order to simplify the integration, we approximate $R_{\rm coll}$ as 
\begin{equation} 
R_{\rm coll} = \left\{
\begin{array}{ll}
 2 r_{\rm e}^2 R_{\rm eject} / r_{\rm p}^2 & {\rm for}~ t \le t_{*}, \\
 2 r_{\rm e}^2 R_{\rm eject} / r_{\rm p}^2 \left( t / t_{*} \right)^3 & {\rm for}~ t > t_{*} .
\label{eq:approximation}
\end{array}
\right.  
\end{equation}
Using above approximation, we obtain $P_{\rm coll} (t_1 , t_2)$ as
\begin{equation} 
P_{\rm coll} (t_1, t_2) = \left\{
\begin{array}{ll}
2 r_{\rm e}^2 R_{\rm eject} / r_{\rm p}^2  \left(t_2 - t_1\right) & {\rm when}~ t_1 \le t_2 \le t_* , \\
r_{\rm e}^2 R_{\rm eject} / r_{\rm p}^2 \left[ 3 t_{*} - 2 t_1 - \left( t_2 / t_{*} \right)^{-3} t_2 \right] & {\rm when}~ t_1 < t_{*} < t_2, \\
r_{\rm e}^2 R_{\rm eject} t_{*}^3 / r_{\rm p}^2 \left( t_1^{-2} - t_2^{-2} \right) & {\rm when}~ t_{*} \le t_1 \le t_2. 
\label{eq:collision_frequency_int}
\end{array}
\right.
\end{equation}

In this paper, we count all collisions that expected to occur behind the parent. Finally, we obtain the probability of collisions experienced by each ejector in one disruption event (collision probability) as
\begin{equation}
P_{\rm coll} \equiv P_{\rm coll} \left( 0, \infty \right)
= \frac{ 3 r_{\rm e}^2 R_{\rm eject} t_{*} }{ r_{\rm p}^2 } 
= \left( \frac{ 72 \rho_{\rm mat} r_{\rm e}^7 R_{\rm eject}^3 }{ r_{\rm p}^4 \Delta v p_{\rm fm} } \right)^{1/3}. 
\label{eq:collision_frequency}
\end{equation}
In addition, we find that $P_{\rm coll} (0, t_*) / P_{\rm coll} = 2/3$. It suggests that most ($67\%$ of) collisions occur within $t_*$, therefore, the time $t_*$ can be considered as the typical timescale of the mutual collision.\footnote{There is an exact solution of the integration. The exact solution of $P_{\rm coll} (0, \infty)$ is smaller than that with the approximation of Eq. (\ref{eq:approximation}) by about $19\%$. In addition, the ratio $P_{\rm coll} (0, t_*) / P_{\rm coll}$ is about $69\%$, instead of $2/3 = 67\%$. }

\subsection{Velocity Dispersion and Ejection Rate}
In order to estimate $\Delta v$ and $R_{\rm eject}$ in Eq. (\ref{eq:collision_frequency}), we consider the hydrodynamics of molten parent particle exposed to gas flow. Before melting, since the cm-sized parent is too large to homogenize internal temperature due to the thermal conduction, the temperature is higher at the parent surface facing to the gas flow than at the center (Yasuda \& Nakamoto 2005, 2006). It causes to form liquid layer at the parent surface. Kato et al. (2006) obtained the internal velocity of the liquid layer by analytically solving the steady hydrodynamics equations for a core-mantle structure with a linear approximation. According to their results, we can approximate a maximum tangential velocity of the liquid layer as $v_{\rm max} \simeq 0.112 \, p_{\rm fm} h / \mu$, where $h$ is the width of the liquid layer and $\mu$ is the viscosity of molten dust particle (see Appendix \ref{sec:kato2006}). Considering that $\Delta v$ is about the same order of magnitude of $v_{\rm max}$, we obtain
\begin{equation}
\Delta v \sim v_{\rm max} \sim p_{\rm fm} h / 10 \mu .
\label{eq:velocity_dispersion}
\end{equation}
The total volume of liquid layer can be roughly estimated as $\sim h r_{\rm p}^2$. When the whole part of the liquid layer fragments into ejectors, the total number of ejectors is $N_{\rm e} \sim h r_{\rm p}^2 / r_{\rm e}^3$. Since it is considered that the fragmentation proceeds in about a fluid crossing time 
\begin{equation}
t_{\rm cross} \sim r_{\rm p} / v_{\rm max} \sim 10 \mu r_{\rm p} / p_{\rm fm} h, 
\label{eq:crossing_time}
\end{equation}
we obtain the ejection rate as
\begin{equation}
R_{\rm eject} \sim N_{\rm e} / t_{\rm cross} \sim r_{\rm p} p_{\rm fm} h^2 / 10 \mu r_{\rm e}^3 .
\label{eq:ejection_rate}
\end{equation}

\subsection{Width of Liquid Layer}
\label{sec:width}
The liquid layer should be thick enough to cause disruption. Kadono \& Arakawa (2005) carried out aerodynamic experiments in which a liquid layer was attached to solid cores, and the breakup of this layer occurred by means of the interaction with a high-velocity gas flow. They discussed that the breakup did not occur at the Weber number defined by $W_e' \equiv p_{\rm fm} h / \gamma_{\rm s}$, where $\gamma_{\rm s}$ is the surface tension, less than $10 - 20$. This result is similar to the finding that the threshold of breakup of liquid droplets without solid cores is $W_e \sim 10$ at $O_h < 0.1$, where $W_e \equiv p_{\rm fm} r_{\rm p} / \gamma_{\rm s}$ is the Weber number for a completely-molten particle and $O_h \equiv \mu / \sqrt{\rho_{\rm mat} r_{\rm p} \gamma_{\rm s}}$ is the Ohnesorge number (e.g., Fig. 1 of Hsiang and Faeth 1995). Based on their results, we consider that ejection will occur at the time when $W_e' = 10$. Therefore, the width of liquid layer is given by
\begin{equation}
h \sim 10 \gamma_{\rm s} / p_{\rm fm}. 
\label{eq:layer_width}
\end{equation}
It should be noted that $h$ does not depend on $r_{\rm p}$. Strictly speaking, we should write $h$ as $h \sim {\rm min} \left[ 10 \gamma_{\rm s} / p_{\rm fm} , r_{\rm p} \right]$ because $h$ cannot exceed $r_{\rm p}$. However, we do not consider the case in which $h > r_{\rm p}$ in this paper, so we simply use Eq. (\ref{eq:layer_width}). 

%The value of $h$ can be larger than $r_{\rm p}$ for small $p_{\rm fm}$ in Eq. (\ref{eq:layer_width}), however, it indicates that the parent particle is not disrupted even when it completely melts. In this study, we consider that there is no disruption in the case of $h / r_{\rm p} = 10 \gamma_{\rm s} / p_{\rm fm} r_{\rm p} > 1$. 

\subsection{Physical Parameters}
\label{sec:physical_parameters}
The parent particle is assumed to be mainly composed of forsterite. The physical parameters adopted in this paper are $\rho_{\rm mat} = 3 \, {\rm g \, cm^{-3}}$ and $\gamma_{\rm s} = 400 \, {\rm dyne \, cm^{-1}}$ (Murase \& McBirney 1973). For other physical parameters, we adopt $r_{\rm p} = 1 \, {\rm cm}$, $r_{\rm e} = 300 \, {\rm \mu m}$, $p_{\rm fm} = 3\times10^4 \, {\rm dyne \, cm^{-2}}$, and $\mu = 10^2 \, {\rm g \, cm^{-1} \, s^{-1}}$ as a standard set of parameters. We also show results for other sets of parameters in \S \ref{sec:parameter_dependence}.

\section{Estimation of Collision Probability}
\label{sec:predicted_collision_frequency}
Substituting Eqs. (\ref{eq:velocity_dispersion}), (\ref{eq:ejection_rate}), and (\ref{eq:layer_width}) to Eq. (\ref{eq:collision_frequency}), we obtain the expression of the collision probability as
\begin{equation}
P_{\rm coll} \sim 
\left( \frac{ 7.2\times10^4 \rho_{\rm mat} \gamma_{\rm s}^5 }{ r_{\rm p} r_{\rm e}^2 p_{\rm fm}^4 \mu^2 } \right)^{1/3} .
\label{eq:collision_frequency2}
\end{equation}
Substituting a standard set of parameters (see \S \ref{sec:physical_parameters}), we obtain $h \sim 1.3 \, {\rm mm}$, $\Delta v \sim 4.0 \, {\rm cm \, s^{-1}}$, and $R_{\rm eject} \sim 2.0\times10^4 \, {\rm s^{-1}}$. From Eq. (\ref{eq:number_density}), we obtain the number density of ejectors behind the parent as $n_{\rm e} \sim 796 \, {\rm cm^{-3}}$ for $t \ll t_{*}$, where $t_{*} \sim 0.013 \, {\rm s}$ (see \S \ref{sec:collision_frequency}). The total volume of these ejectors is given by $(4/3) \pi r_{\rm e}^3 n_{\rm e} \sim 0.09$, so they occupy about $9\%$ of the space behind the parent in volume. Finally, the collision probability is $P_{\rm coll} \sim 0.67$. Surprisingly, above estimation is larger than the observed fraction of compound chondrules by one order of magnitude or more. If we assume that all collisions lead to compound chondrule formation, this result suggests that the fragment-collision model can account for the observed fraction when only $\sim 10\%$ or less of all chondrules formed via the fragmentation events of cm-sized parent dust particles (also see \S \ref{sec:observational_fraction}). Since the probability is close to unity, most of ejectors would experience one mutual collision at least one time in average. In addition, some of them might experience multiple collisions (more than twice) and form multiple compound chondrules.

\subsection{Parameter Dependence}
\label{sec:parameter_dependence}
The expression of collision probability, Eq. (\ref{eq:collision_frequency2}), is complex to understand how each parameter affects the result. In order to see in details, we explicitly write the dependences of physical parameters as follows:
\begin{eqnarray}
h &\propto& p_{\rm fm}^{-1} , \label{eq:layer_width11} \\
N_{\rm e} &\propto& r_{\rm p}^2 r_{\rm e}^{-3} p_{\rm fm}^{-1} , \label{eq:ejector_number11} \\
t_{\rm cross} &\propto& r_{\rm p} \mu , \label{eq:crossing_time11} \\
R_{\rm eject} &\propto& r_{\rm p} r_{\rm e}^{-3} p_{\rm fm}^{-1} \mu^{-1} , \label{eq:ejection_rate11} \\
\delta V_0 &\propto& r_{\rm p}^2 \mu^{-1} , \label{eq:volume0_11} \\
n_{\rm e} &\propto& r_{\rm p}^{-1} r_{\rm e}^{-3} p_{\rm fm}^{-1} , \label{eq:number_density11} \\
\sigma_{\rm coll} &\propto& r_{\rm e}^2 , \label{eq:cross_section11} \\
\Delta v &\propto& \mu^{-1} , \label{eq:velocity_dispersion11} \\
R_{\rm coll} &\propto& r_{\rm p}^{-1} r_{\rm e}^{-1} p_{\rm fm}^{-1} \mu^{-1} , \label{eq:collision_rate11} \\
t_{*} &\propto& r_{\rm p}^{2/3} r_{\rm e}^{1/3} p_{\rm fm}^{-1/3} \mu^{1/3} , \label{eq:t_star11} \\
P_{\rm coll} &\propto& r_{\rm p}^{-1/3} r_{\rm e}^{-2/3} p_{\rm fm}^{-4/3} \mu^{-2/3} . \label{eq:collision_frequency11}
\end{eqnarray}
Let us see the dependence on the ejector radius $r_{\rm e}$ at first. The total number of ejectors contained in the liquid layer, $N_{\rm e}$, decreases with increase of $r_{\rm e}$ as $\propto r_{\rm e}^{-3}$. The ejection timescale $t_{\rm cross}$ does not depend on $r_{\rm e}$. Therefore, the ejection rate $R_{\rm coll} \sim N_{\rm e} / t_{\rm cross}$ decreases as $r_{\rm e}$ increases. We consider only the collision at the region close to the parent particle because most ($67\%$ of) collisions occur at the early phase of $t \le t_{*}$ (see \S \ref{sec:collision_probability}). In this case, since the volume $\delta V_0$ does not depend on $r_{\rm e}$, the number density of ejectors $n_{\rm e} \propto R_{\rm eject} / \delta V_0$ decreases with increase of $r_{\rm e}$ as $\propto r_{\rm e}^{-3}$. The collisional cross section $\sigma_{\rm coll}$ is proportional to $r_{\rm e}^2$ and the velocity dispersion $\Delta v$ does not depend on $r_{\rm e}$, so the collision rate $R_{\rm coll} \sim \sigma_{\rm coll} n_{\rm e} \Delta v$ is inversely proportional to $r_{\rm e}$. We can roughly estimate the collision probability by $P_{\rm coll} \sim R_{\rm coll} t_{*}$, which is smaller than Eq. (\ref{eq:collision_frequency}) by a factor of 2/3, and $t_{*}$ is proportional to $r_{\rm e}^{1/3}$, so it is found that $P_{\rm coll}$ decreases with increase of $r_{\rm e}$.

We will show other dependences of $P_{\rm coll}$ on the parent radius $r_{\rm p}$, the gas ram pressure $p_{\rm fm}$, and viscosity of molten parent particle $\mu$ as below.

\subsubsection{Parent Radius}
\begin{figure}[p]
\center
\includegraphics[scale=.7, angle=-90.]{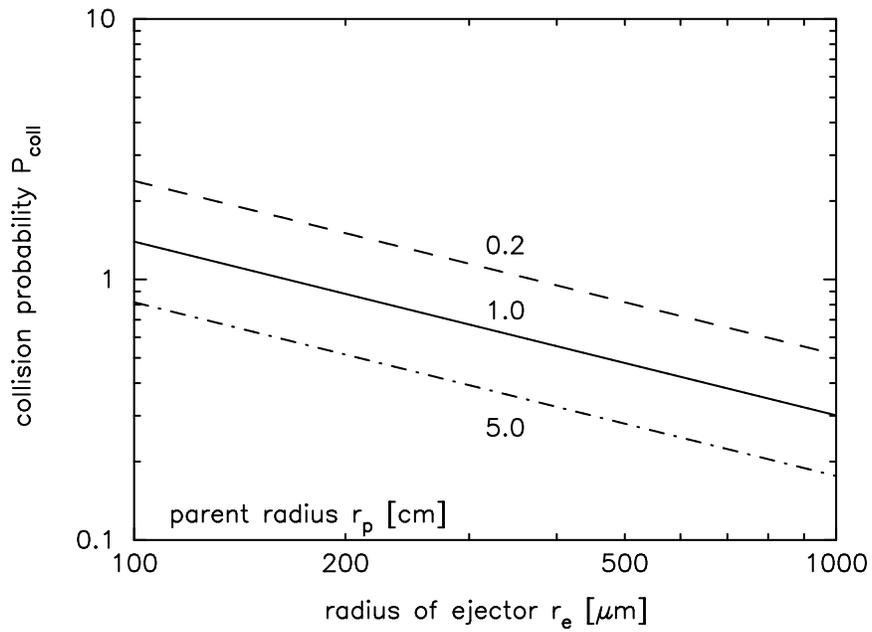}
\caption{Collision probability $P_{\rm coll}$ are plotted as a function of ejector radius $r_{\rm e}$ for various parent radii of $r_{\rm p} = 0.2$, $1.0$, and $5.0 \, {\rm cm}$. We adopt standard values for other parameters, $p_{\rm fm} = 3\times10^4 \, {\rm dyne \, cm^{-2}}$ and $\mu = 100 \, {\rm g \, cm^{-1} \, s^{-1}}$ (see \S \ref{sec:physical_parameters}). }
\label{fig:plot_noshadow_rp}
\end{figure}

Fig. \ref{fig:plot_noshadow_rp} shows the value of $P_{\rm coll}$ as a function of $r_{\rm e}$ for various values of $r_{\rm p} = 0.2$ (dashed), $1.0$ (solid), and $5.0 \, {\rm cm}$ (dotted-dashed), respectively. It is found that $P_{\rm coll}$ is larger for smaller parent than for larger one as estimated by Eq. (\ref{eq:collision_frequency11}). The disruption timescale is longer for larger $r_{\rm p}$, on the other hand, $N_{\rm e}$ increases with increase of $r_{\rm p}$ more steeply, so $R_{\rm eject}$ increases as $r_{\rm p}$ increases. However, for larger $r_{\rm p}$, ejectors are disrupted into more wider region. As a result, $n_{\rm e}$ and $R_{\rm coll}$ are inversely proportional to $r_{\rm p}$. The time $t_{*}$ increases with increase of $r_{\rm p}$, but it cannot completely cancel the dependence of $R_{\rm coll}$. Finally, we obtain $P_{\rm coll} \propto r_{\rm p}^{-1/3}$ as seen in Eq. (\ref{eq:collision_frequency11}). For the parameter range we adopted in Fig. \ref{fig:plot_noshadow_rp}, $P_{\rm coll}$ ranges about from $0.2$ to $2$.

\hspace{1cm} {\bf [Figure 2]}

\subsubsection{Gas Ram Pressure}
\begin{figure}[p]
\center
\includegraphics[scale=.7, angle=-90.]{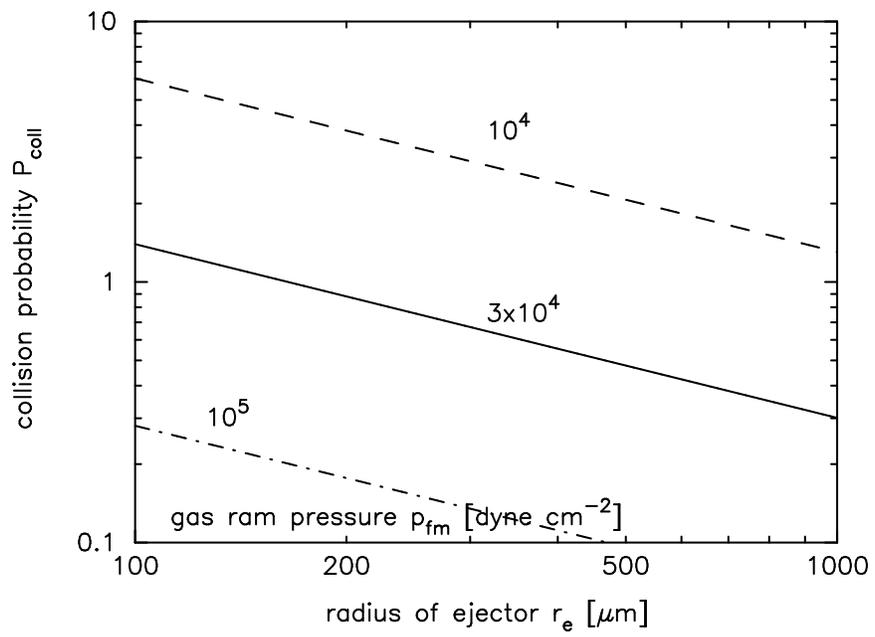}
\caption{Same as Fig. \ref{fig:plot_noshadow_rp} except for various gas ram pressures of $p_{\rm fm} = 10^4$, $3\times10^4$, and $10^5 \, {\rm dyne \, cm^{-2}}$. Other parameters are described in \S \ref{sec:physical_parameters}. }
\label{fig:plot_noshadow_pf}
\end{figure}

Fig. \ref{fig:plot_noshadow_pf} shows the value of $P_{\rm coll}$ as a function of $r_{\rm e}$ for various values of $p_{\rm fm} = 10^4$ (dashed), $3\times10^4$ (solid), and $10^5 \, {\rm dyne \, cm^{-2}}$ (dotted-dashed), respectively. It is found that $P_{\rm coll}$ is larger for weaker gas ram pressure than for stronger one as estimated by Eq. (\ref{eq:collision_frequency11}). The stronger gas flow can disrupt thiner liquid layer (Eq. \ref{eq:layer_width11}), so $N_{\rm e}$ and $R_{\rm eject}$ decreases as $p_{\rm fm}$ increases. This results in that $n_{\rm e}$ and $R_{\rm coll}$ are also inversely proportional to $p_{\rm fm}$. In addition, the strong gas flow scatters disrupted ejectors rapidly, so $t_{*}$ decreases as $p_{\rm fm}$ increases. Finally, we obtain $P_{\rm coll} \propto p_{\rm fm}^{-4/3}$ as seen in Eq. (\ref{eq:collision_frequency11}). For the parameter range we adopted in Fig. \ref{fig:plot_noshadow_pf}, $P_{\rm coll}$ ranges about from $0.06$ to $6$.

\hspace{1cm} {\bf [Figure 3]}

\subsubsection{Viscosity}
\begin{figure}[p]
\center
\includegraphics[scale=.7, angle=-90.]{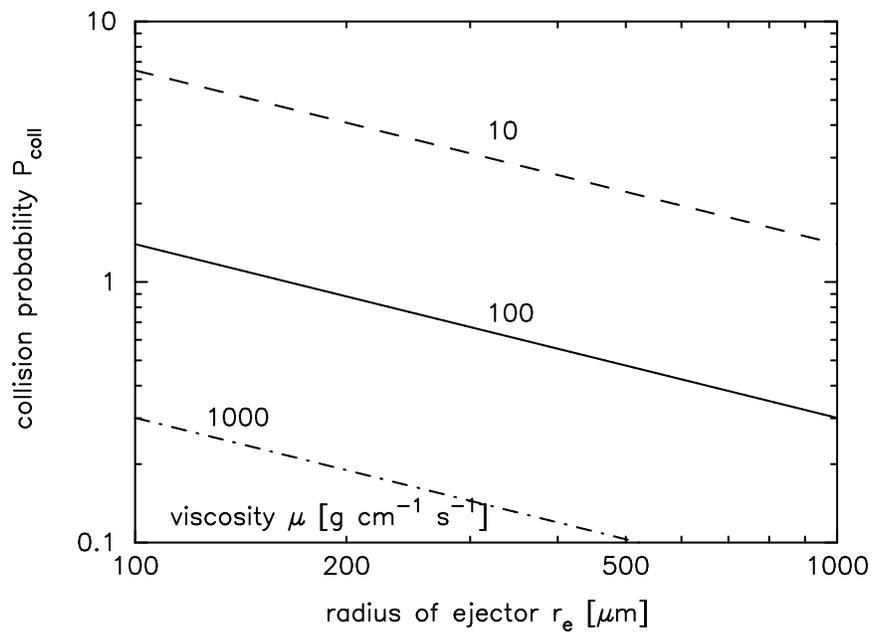}
\caption{Same as Fig. \ref{fig:plot_noshadow_rp} except for various viscosities of $\mu = 10$, $10^2$, and $10^3 \, {\rm g \, cm^{-1} \, s^{-1}}$. Other parameters are described in \S \ref{sec:physical_parameters}.}
\label{fig:plot_noshadow_mu}
\end{figure}

Fig. \ref{fig:plot_noshadow_mu} shows the value of $P_{\rm coll}$ as a function of $r_{\rm e}$ for various values of $\mu = 10$ (dashed), $10^2$ (solid), and $10^3 \, {\rm g \, cm^{-1} \, s^{-1}}$ (dotted-dashed), respectively. It is found that $P_{\rm coll}$ increases as $\mu$ decreases as estimated by Eq. (\ref{eq:collision_frequency11}). In a highly viscous case, $R_{\rm eject}$ becomes small because of a long ejection timescale (Eq. \ref{eq:ejection_rate11}). However, in this case, since the velocity dispersion $\Delta v$ also decreases and ejectors are disrupted into narrow region, $n_{\rm e}$ does not depend on $\mu$. Therefore, $R_{\rm coll}$ is inversely proportional to $\mu$, which is due to the dependence of $\Delta v$. Considering the dependence of $t_{*}$, we obtain $P_{\rm coll} \propto \mu^{-2/3}$ as seen in Eq. (\ref{eq:collision_frequency11}). For the parameter range we adopted in Fig. \ref{fig:plot_noshadow_mu}, $P_{\rm coll}$ ranges about from $0.06$ to $6$.

\hspace{1cm} {\bf [Figure 4]}

\section{Parameter Range}
\label{sec:parameter_range}

\subsection{Parent Radius}
\label{sec:parent_radius}
In Eq. (\ref{eq:layer_width}), we estimated the layer width $h$ from the disruption condition of partially-molten parent particle. However, it should be noted that this estimation is not valid for much larger parent particle because the disruption timescale becomes larger than the thermal conduction timescale. In this case, the layer width $h$ will increase by the rapid thermal conduction before disruption as explained below. The thermal conduction timescale in the liquid layer $t_{\rm cond}$ is given by $t_{\rm cond} \sim \rho_{\rm mat} C h^2 / \kappa \sim 100 \rho_{\rm mat} C \gamma_{\rm s}^2 / \kappa p_{\rm fm}^2$, in the second sign of similarity we substitute Eq. (\ref{eq:layer_width}), where $C$ is the specific heat per unit mass and $\kappa$ is the heat conductivity. On the other hand, we obtain the disruption timescale $t_{\rm cross} \sim r_{\rm p} \mu / \gamma_{\rm s}$ by substituting Eq. (\ref{eq:layer_width}) to Eq. (\ref{eq:crossing_time}). The layer width $h$ would not change significantly during disruption if the condition of $t_{\rm cross} < t_{\rm cond}$ is satisfied. We rewrite this condition as
\begin{equation}
r_{\rm p} \la \frac{ 100 \rho_{\rm mat} C \gamma_{\rm s}^3 }{ \mu \kappa p_{\rm fm}^2 } .
\label{eq:rp_crit}
\end{equation}
Adopting $C = 10^7 \, {\rm erg \, g^{-1} \, K^{-1}}$ and $\kappa = 4\times10^5 \, {\rm erg \, cm^{-1} \, s^{-1} \, K^{-1}}$ (Murase \& McBirney 1973), we find $r_{\rm p} \la 5.3 \, {\rm cm}$ for a standard set of parameters (see \S \ref{sec:physical_parameters}). If $r_{\rm p}$ is much larger than this value, the layer width $h$ changes during disruption because the thermal conduction can rapidly transfer energy inside the parent and melt deeply. In this study, we do not treat such situation because we consider cm-sized parent particle, which satisfies the condition of Eq. (\ref{eq:rp_crit}).

\subsection{Gas Ram Pressure}
The gas ram pressure $p_{\rm fm}$ should be sufficient to disrupt the molten parent particle. For simplicity, we consider a completely-molten parent particle here ($h = r_{\rm p}$). The disruption will occur when the Weber number $W_e \equiv p_{\rm fm} r_{\rm p} / \gamma_{\rm s}$ exceeds about 10 (Hsiang \& Faeth 1995). Therefore, the condition that $p_{\rm fm}$ should satisfy is given by 
\begin{equation}
p_{\rm fm} \ga 10 \gamma_{\rm s} / r_{\rm p} .
\label{eq:ram_pressure}
\end{equation}
We find $p_{\rm fm} \ga 4\times10^3 \, {\rm dyne \, cm^{-2}}$ for a standard set of parameters (see \S \ref{sec:physical_parameters}). 

\begin{figure}[p]
\center
\includegraphics[scale=.9, angle=0.]{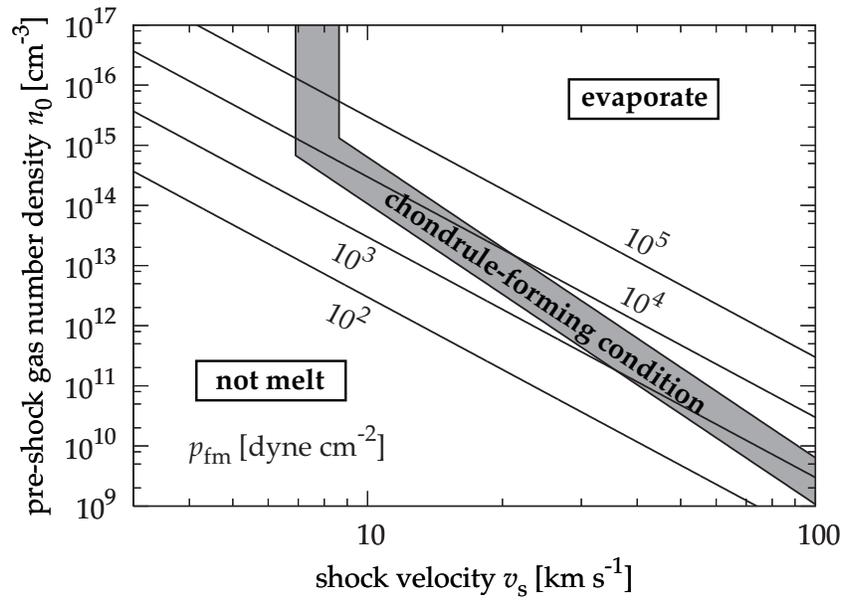}
\caption{Gas ram pressure affecting the molten parent particle behind the shock front. The horizontal axis is the shock velocity $v_{\rm s}$ and the vertical axis is the pre-shock gas number density $n_0$. The gray region is the chondrule-forming shock condition in which the gas frictional heating is sufficient to melt the precursor dust particle but not so strong as it evaporates the dust completely (Iida et al. 2001).}
\label{fig:shock_condition}
\end{figure}

Fig. \ref{fig:shock_condition} shows the gas ram pressure $p_{\rm fm}$ expected to affect the molten parent particle just behind the shock front (solid lines) as a function of the shock velocity $v_{\rm s}$ and the pre-shock gas number density $n_0$. The gray region is the chondrule-forming shock condition in which the gas frictional heating is sufficient to melt the precursor dust particle but not so strong as it evaporates the dust completely (Iida et al. 2001). The gas ram pressure is $p_{\rm fm} \sim 10^4 - 10^5 \, {\rm dyne \, cm^{-2}}$ when $n_0 \sim 10^{15} - 10^{16} \, {\rm cm^{-3}}$ and $v_{\rm s} \sim 7 - 10 \, {\rm km \, s^{-1}}$. The shocks associated with gravitational instability (e.g., Boss \& Durisen 2005) or planetesimal bow shocks (e.g., Hood 1998, Weidenschilling et al. 1998) might be preferable for such situation.

\hspace{1cm} {\bf [Figure 5]}

\subsection{Viscosity}
The viscosity of molten parent particle should be small enough to cause disruption. The timescale of disruption is given by $t_{\rm cross} \sim r_{\rm p} \mu / \gamma_{\rm s}$ (see Eqs. \ref{eq:crossing_time} and \ref{eq:layer_width}). On the other hand, the heating event ceases within the timescale that the parent dust particle stops against the ambient gas (Iida et al. 2001). The timescale is the stopping time, which is given by $t_{\rm stop} \sim \rho_{\rm mat} r_{\rm p} v_{\rm g} / p_{\rm fm}$, where $v_{\rm g}$ is the relative velocity between ambient gas and dust particle.\footnote{Substituting $p_{\rm fm} = \rho_{\rm g} v_{\rm g}^2$, where $\rho_{\rm g}$ is the gas density, we obtain a well-known expression of the stopping time $t_{\rm stop} \sim \rho_{\rm mat} r_{\rm p} / \rho_{\rm g} v_{\rm g}$.} If $t_{\rm cross} > t_{\rm stop}$, the gas drag heating ceases before the disruption occurs. Therefore, the viscosity at the disruption satisfies the condition of $t_{\rm cross} < t_{\rm stop}$. We rewrite this condition as
\begin{equation}
\mu \la \frac{ \rho_{\rm mat} \gamma_{\rm s} v_{\rm g} }{ p_{\rm fm} } .
\label{eq:viscosity}
\end{equation}
Substituting $v_{\rm g} = v_{\rm s} = 8 \, {\rm km \, s^{-1}}$ (see Fig. \ref{fig:shock_condition}), we find $\mu \la 3.2\times10^4 \, {\rm g \, cm^{-1} \, s^{-1}}$ for a standard set of parameters (see \S \ref{sec:physical_parameters}). 

In addition, the viscosity should be large enough to adopt the steady solutions (Kato et al. 2006) as discussed below. The liquid layer will be ejected within the timescale of $t_{\rm cross} \sim r_{\rm p} \mu / \gamma_{\rm s}$ (see Eqs. \ref{eq:crossing_time} and \ref{eq:layer_width}), and it decreases as $\mu$ decreases. For example, we obtain $t_{\rm cross} \sim 2.5\times10^{-4} \, {\rm s}$ for $r_{\rm p} = 1 \, {\rm cm}$ and $\mu = 0.1 \, {\rm g \, cm^{-1} \, s^{-1}}$. However, the liquid layer cannot be accelerated to the velocity given by Eq. (\ref{eq:velocity_dispersion}) in such a short duration. Considering that the momentum of the gas flow converts to the motion of the liquid layer effectively, the timescale in which the liquid layer is accelerated to $\Delta v$ is given by $t_{\rm acc} \sim m \Delta v / F$, where the mass of the liquid layer is $m \sim 2 \pi \rho_{\rm mat} r_{\rm p}^2 h$, the force due to the gas flow is $F = \pi r_{\rm p}^2 p_{\rm fm}$, and the velocity dispersion is $\Delta v \sim \gamma_{\rm s} / \mu$ (see Eqs. \ref{eq:velocity_dispersion} and \ref{eq:layer_width}). Then we obtain
\begin{equation}
t_{\rm acc} \sim 20 \rho_{\rm mat} \gamma_{\rm s}^2 / p_{\rm fm}^2 \mu .
\label{eq:acceleration_time}
\end{equation}
When $t_{\rm acc} > t_{\rm cross}$, the liquid layer does not reach the steady solutions before ejection. Therefore, the condition of $t_{\rm acc} < t_{\rm cross}$ should be satisfied in our model. This condition can be rewritten as
\begin{equation}
\mu \ga \left( \frac{ 20 \rho_{\rm mat} \gamma_{\rm s}^3 }{ r_{\rm p} p_{\rm fm}^2 } \right)^{1/2} .
\label{eq:viscosity2}
\end{equation}
We find $\mu \ga 2 \, {\rm g \, cm^{-1} \, s^{-1}}$ for a standard set of parameters (see \S \ref{sec:physical_parameters}).

\section{Comparison with Observations}
\label{sec:observation}

%\subsection{Timescale of Collisions}
%\label{sec:timescale_of_collisions}
%Deleted by Authors (Oct. 10, 2007). 
%Collisions between molten ejectors should occur before they solidify to be so stiff at their surfaces. Since most ($67\%$ of) collisions would be done by $t = t_{*}$, the typical timescale of collisions is given by $t_*$ (see Appendix \ref{sec:integration}). In a standard set of parameters, we obtain $t_{*} \simeq 0.01 \, {\rm s}$. On the other hand, the fastest cooling mechanism of the molten ejectors is the radiative cooling. The cooling rate is given by
%\begin{equation}
%R_{\rm cool} = \frac{ 3 \epsilon_{\rm emit} \sigma_{\rm SB} T_{\rm e}^4 }{ r_{\rm e} \rho_{\rm mat} C }, 
%\label{eq:radiative_cooling}
%\end{equation}
%where $\epsilon_{\rm emit}$ is the emission coefficient, $\sigma_{\rm SB}$ is the Stefan-Boltzmann constant, and $T_{\rm e}$ is the temperature of the fragment. We obtain a typical value of $R_{\rm cool} \simeq 10^3 \, {\rm K \, sec^{-1}}$ for $\epsilon_{\rm emit} = 1$ and $T_{\rm e} = 1600 \, {\rm K}$. Adopting this cooling rate, the timescale for ejectors to cool by $100 \, {\rm K}$ is $t_{\rm cool} \sim 0.08 \, {\rm s}$. Therefore, most collisions can occur before the molten ejectors cool significantly. 

\subsection{Fraction of Compound Chondrules}
\label{sec:observational_fraction}
We found that the collision probability estimated by our model, $P_{\rm coll}$, is close to unity, which is larger than the observed fraction of compound chondrules by one order of magnitude or more (see \S \ref{sec:predicted_collision_frequency}). If we assume that all collisions lead to compound chondrule formation, this result suggests that the fragment-collision model can account for the observed fraction when about $10\%$ or less of all chondrules formed via the fragmentation events. The fraction of chondrules which have undergone the fragmentation event depends on a size distribution of precursor dust particles in chondrule-forming region. Nomura et al. (2007) solved coagulation equations for various sizes of settling dust particles in the solar nebula. They obtained the dust size distributions for various positions in the nebula and ages of the nebula. According to their results, for example, the dust particles with the radii from $1 \, {\rm mm}$ to $1 \, {\rm cm}$ have the size distribution similar to $dn \propto r^{-4} d(\log r)$, where $r$ is the dust radius, in regions near midplane at the distance of $1 \, {\rm AU}$ or $10 \, {\rm AU}$ and at the age of $10^6 \, {\rm yr}$ (it corresponds to the line of the slope $-1$ in Fig. 4 of their paper). This suggests that there are $10^4$ of mm-sized dust particles per a cm-sized one. Assuming that the cm-sized particle is disrupted into mm-sized ejectors in a shock-wave heating,\footnote{In the shock-wave heating model, all of the equations governing the evolution of the precursor dust particle are scaled with the initial precursor radius $r_0$ as long as the post-shock gas properties (temperature, density, and so forth) are spatially uniform (see \S 4.3 in Miura et al. 2002, Icarus 160, 258). It means that the peak temperature of the precursor dust particle does not depend on the radius. The reason of the scaling law is that all of the heating rates (due to the gas friction, radiation, and so forth) and cooling rates (due to the radiative cooling, latent heat of evaporation, and so forth) are given as the rates per unit area and the peak temperature is determined by the balance between heating and cooling, so the dependence on the particle radius does not appear. Namely, if mm-sized dust particles melt (and chondrules are formed), cm-sized dust particles also melt in the same shock condition.} total number of ejectors is estimated as $\sim 10^3$. If we also assume that all of them become compound chondrules with two constituents, about 500 sets of compounds are expected to form. In this case, the fraction of compound chondrules is $\sim 5\%$ ($ = 500 / 10^4$). This is very close to the observed fraction of compound chondrules. 

However, all collisions do not necessarily lead to compound chondrule formation if we consider the situation of mutual collision in more detail. We discuss the efficiency of compound chondrule formation in \S \ref{sec:efficiency}.

\subsection{Oxygen Isotopic Composition}
In the fragment-collision model we proposed in this paper, the constituent chondrules of compounds are likely to have similar compositions because ejectors originate from the same parent. Akaki \& Nakamura (2005) measured the oxygen isotopic compositions for 3 sets of blurred-type compounds, 6 sets of adhering- or consorting-type compounds, and 2 sets of enveloping-type compounds. It was found that in a three-isotopic diagram, all sets of blurred-, adhering-, and consorting-types fall in the typical range obtained for single chondrules from the same CV3 chondrites. These results suggest that the two constituent chondrules of these compounds originated from the same dust reservoirs as those single chondrules. These observations are consistent with our new model for compound chondrule formation. 

In contrast, in one set of the enveloping-types, the oxygen isotopic compositions differ between two constituent chondrules. This result might suggest that this enveloping-type compound has not formed by our model but by the relict grain model, in which fine-grained dust particles accreted on the surface of already-formed primary were melted in a second heating event (Wasson 1993, Wasson et al. 1995).

\subsection{Textural Types}
There is a dependence of observed fraction of compound chondrules on textural types of component chondrules. Gooding \& Keil (1981) found that compound chondrules with non-porphyritic pairs are more frequent than that with porphyritic pairs. From their thin section results, they estimated that 13\% of non-porphyritic and 2\% of porphyritic chondrules are compound or cratered, which are interpreted as products of collisions between plastic chondrules. Akaki \& Nakamura (2005) reported the same tendency. It is considered that porphyritic and non-porphyritic textures have been formed from partially- and completely-molten dust particles, respectively (Lofgren \& Russell 1986). It is also naturally considered that a partially-molten dust particle is highly-viscous, in contrast, completely-molten one has lower viscosity. Based on above assumption, the dependence of the observed fraction of compound chondrules on the textural types might reflect the dependence of the collision probability on the viscosity of molten parent dust particle. Actually, our model predicts that lower viscosity results into larger collision probability as shown in Fig. \ref{fig:plot_noshadow_mu}. This result is consistent with the observations. 

In addition, various types of textures are seen in compound chondrules, e.g., P-P, P-NP, and NP-NP pairs, where P and NP stand for porphyritic and non-porphyritic textures, respectively (Wasson et al. 1995). In contrast, our model seems not to account for compound chondrules with porphyritic textures because the following properties are implied: (a) completely-molten ejectors are extracted from completely-molten parent particle, and (b) temperatures of all ejectors are the same at the extraction. However, regarding the point (a), it can be considered that the partially-molten parent particle, which includes tiny unmelted cores inside, behaves as a fluid if the molten part occupies the most volume of the parent. In this case, the partially-molten ejectors could be extracted from the partially-molten parent and the porphyritic textures might be formed. Regarding the point (b), we can consider the case that the temperatures of ejectors are not uniform at the extraction. If the initial temperatures of ejectors are different, various types of textures can be formed in compound chondrules. In order to verify above hypotheses, we must quantitatively investigate the thermal evolutions of the parent dust particle and each ejector at disruption. It requires to carry out the three-dimensional (thermo-)hydrodynamics simulation and/or aerodynamical disruption experiment.

In addition, Connolly \& Hewins (1995) reported that porphyritic textures can be reproduced from totally-molten droplets by dust seeding. Based on their results, even if ejectors extracted from the parent are totally-molten, they can obtain porphyritic textures inside them with the help of the dust seeding. Considering that such seeded ejectors collide with others before solidification, compound chondrules including porphyritic textures might be formed. Therefore, the effect of the dust seeding in compound chondrule formation is also an important issue that should be investigated in the future.

\subsection{Efficiency of Compound Chondrule Formation}
\label{sec:efficiency}
In \S \ref{sec:observational_fraction}, we implicitly assumed that all collisions between ejectors make compound chondrules, however, it is an upper limit of the probability of compound chondrule formation: for example, a collision between two totally-molten and high-temperature (low-viscous) ejectors does not lead to a compound chondrule because they will fuse into a single droplet. In addition, a collision after significant cooling has a difficulty in compound chondrule formation because both ejectors will solidify completely. Therefore, we have to clarify the collision probability during ejectors are in the moderate cooling phase. 

Assuming that ejectors cool due to the radiative cooling, the cooling rate is given by
\begin{equation}
R_{\rm cool} = \frac{ 3 \sigma_{\rm SB} \left( \epsilon_{\rm emit} T_{\rm e}^4 - \epsilon_{\rm abs} T_{\rm rad}^4 \right) }{ r_{\rm e} \rho_{\rm mat} C }, 
\label{eq:radiative_cooling}
\end{equation}
where $\epsilon_{\rm emit}$ and $\epsilon_{\rm abs}$ are the emission and absorption coefficient, respectively, $\sigma_{\rm SB}$ is the Stefan-Boltzmann constant, $T_{\rm e}$ is the temperature of the ejectors, $T_{\rm rad}$ is the effective temperature of the ambient radiation field (Desch \& Connolly 2002, Miura \& Nakamoto 2006), and $C$ is the specific heat. The cooling timescale in which the temperature decreases by $\Delta T$ is given by $\sim \Delta T / R_{\rm cool}$. Here, we assume that $\Delta T = 100 \, {\rm K}$ and $300 \, {\rm K}$ are required for moderate cooling and complete solidification, respectively. The effective radiation temperature $T_{\rm rad}$ should be lower than $1273 \, {\rm K}$ in order to prevent the isotopic fractionation of sulfur in chondrules (Miura \& Nakamoto 2006). The cooling rate is reduced by the term of $T_{\rm rad}$, however, it is a minor effect in the cooling phase (if we assume $T_{\rm rad} = 1000 \, {\rm K}$, we obtain $T_{\rm rad}^4 / T_{\rm e}^4 = 0.15$), so we neglect this term for simplicity. Assuming that $\epsilon_{\rm emit} = 1$ and $T_{\rm e} = 1600 \, {\rm K}$, compound chondrules can be formed only in the phase from $t_{\rm cool} = 0.08 \, {\rm s}$ to $t_{\rm solid} = 0.24 \, {\rm s}$ (moderate cooling phase). Since the timescale of the mutual collision ($t_{*} \sim 0.01 \, {\rm s}$, see \S \ref{sec:collision_probability}) is shorter than $t_{\rm cool}$ and $t_{\rm solid}$, we obtain the collision probability between $t_{\rm cool}$ and $t_{\rm solid}$ as (see Eq. \ref{eq:collision_frequency_int})
\begin{equation}
P'_{\rm coll} \equiv P_{\rm coll} \left( t_{\rm cool} , t_{\rm solid} \right) \simeq \frac{ r_{\rm e}^2 R_{\rm eject} t_{*} }{ r_{\rm p}^2 } \left( \frac{ t_{*} }{ t_{\rm cool} } \right)^2 , 
\label{eq:collision_probability_dash}
\end{equation}
where we neglect the term of $t_{\rm solid}$ because $t_{\rm solid}^{-2} \ll t_{\rm cool}^{-2}$. Comparing with Eq. (\ref{eq:collision_frequency}), we find that $P'_{\rm coll}$ is smaller than $P_{\rm coll}$ by a factor of $Q \equiv P'_{\rm coll} / P_{\rm coll} = (t_{*} / t_{\rm cool})^2 / 3 \sim 5 \times 10^{-3}$. The physical meaning of $Q$ is the efficiency of compound chondrule formation per a mutual collision. It means that almost all (99.5\% of) collisions do not lead to compound chondrule formation. Taking into account the low efficiency, the probability of compound chondrule formation is about one order of magnitude smaller than the observed fraction of compound chondrules, although this estimation is larger than that estimated from the random collision model (Gooding \& Keil 1981, Sekiya \& Nakamura 1996). 

However, we can consider other possibilities which enhance the collision probability. One possibility is that the gas flow is blocked by the parent particle and ejectors. In this case, the shadowed region in which there is no effect of the gas flow appears behind them. Since the ejectors in the shadowed region are not accelerated by the gas flow, the concentration of ejectors will be higher than the case neglecting this effect. The second possibility is to consider the collisions between different-sized ejectors. The relative velocity between the different-sized ejectors is given by $\sim \Delta v + | v_{\rm rel} |$, where $| v_{\rm rel} |$ is the bulk relative velocity between different-sized ejectors due to the difference of the acceleration (see \S \ref{sec:collision_frequency}). This large relative velocity between different-sized ejectors is expected to result into large collision probability as long as the collisional disruption does not occur (also see \S \ref{sec:size_ratio}). The third possibility is to consider that the disruption from partially-molten parent particle is possible. In this case, the moderate cooling might not be needed, so $t_{\rm cool}$ can be shorter and it leads to the increase of $P'_{\rm coll}$ (see Eq. \ref{eq:collision_probability_dash}). These possibilities are very important to investigate in detail, however, they are beyond the scope of this paper. We are planning to investigate these issues and the results will appear in the forthcoming papers.

\subsection{Size Ratio of Primary and Secondary}
\label{sec:size_ratio}
Wasson et al. (1995) measured median diameters of primaries and secondaries in compound chondrules and the ratio of the ``diameter" of the secondary divided by that of the primary. They found that the mean and median of the ratio are about 0.3 and 0.25, respectively. These results suggest that compound chondrules with different-sized pairs appear more frequently than that with same-sized pairs. On the other hand, in our model, if there is a size difference between colliding two ejectors moving in the gas flow, it leads to a large relative velocity between them because the acceleration $a$ depends on the ejector radius (see \S \ref{sec:collision_frequency}). The large relative velocity will promote more frequent collisions as commented in \S \ref{sec:efficiency}. In contrast, the undesirably large relative velocity causes disruption of melt droplets upon impact, which is an opposite scenario against compound chondrule formation. Here, we discuss the appropriate range of diameter ratio of the two ejectors that does not exceed the limit of the critical destruction velocity.  

\begin{figure}[p]
\center
\includegraphics[scale=.9, angle=0.]{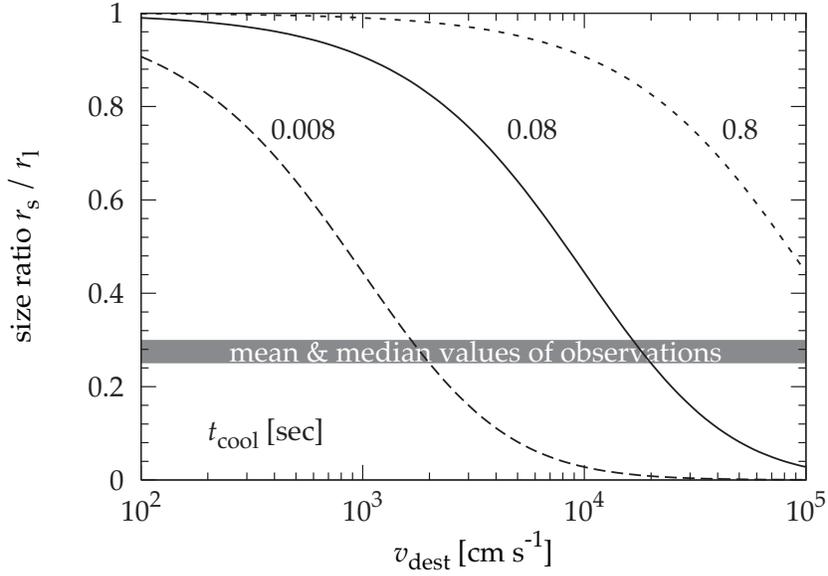}
\caption{Appropriate ranges of size ratio of the small chondrule (radius of $r_{\rm s}$) to large one (radius of $r_{\rm l}$) for compound chondrule formation are displayed as a function of assumed destruction velocity $v_{\rm dest}$ in the case of $r_{\rm l} = 300 \, {\rm \mu m}$ and $p_{\rm fm} = 3 \times 10^4 \, {\rm dyne \, cm^{-2}}$. According to Ciesla (2006), $v_{\rm dest} \sim 10^4 \, {\rm cm \, s^{-1}}$. The solid, dashed, and dotted curves are criteria of the destructive collision for the timescales of the moderate cooling $t_{\rm cool}$ of $0.008 \, {\rm sec}$, $0.08 \, {\rm sec}$, and $0.8 \, {\rm sec}$, respectively, below which the relative velocity between the small and large chondrules is too large for coalescence. The gray region suggests the mean or median values from the observational data (Wasson et al. 1995). }
\label{fig:sizeratio}
\end{figure}

Ejectors are accelerated by the gas flow with the constant acceleration $a = 3 p_{\rm fm} / 4 r_{\rm e} \rho_{\rm mat}$ (see \S 2.1). The velocity relative to the parent particle in the $z$-direction, $v_z$, and the position from the parent, $z$, are given by
\begin{equation}
v_z = a t = \frac{ 3 p_{\rm fm} t }{ 4 r_{\rm e} \rho_{\rm mat} } , 
\label{eq:velocity}
\end{equation}
\begin{equation}
z = \frac{1}{2} a t^2 = \frac{ 3 p_{\rm fm} t^2 }{ 8 r_{\rm e} \rho_{\rm mat} } , 
\label{eq:position}
\end{equation}
where $t$ is the time after extraction. It is considered that the compound chondrule formation occurs after the ejector cools moderately (see \S \ref{sec:efficiency}). Therefore, we consider the region of
\begin{equation}
z > z_{\rm cool} \equiv \frac{ 3 p_{\rm fm} t_{\rm cool}^2 }{ 8 r_{\rm e} \rho_{\rm mat} } .
\label{eq:z_min}
\end{equation}
Next, eliminating $t$ from Eqs. (\ref{eq:velocity}) and (\ref{eq:position}), we obtain
\begin{equation}
v_z = \left( \frac{ 3 p_{\rm fm} z }{ 2 r_{\rm e} \rho_{\rm mat} } \right)^{1/2} .
\end{equation}
Since the velocity $v_z$ depends on the ejector radius $r_{\rm e}$, there is the relative velocity $v_{\rm rel}$ between the large ejector with radius $r_{\rm l}$ and the small ejector with radius $r_{\rm s}$ at the same position. It is given by
\begin{equation}
v_{\rm rel} = \left( \frac{ 3 p_{\rm fm} z }{ 2 r_{\rm l} \rho_{\rm mat} } \right)^{1/2}
\left[ \left( \frac{ r_{\rm l} }{ r_{\rm s} } \right)^{1/2} - 1 \right] .
\label{eq:relative_velocity}
\end{equation}
It is found that $v_{\rm rel}$ increases as $z$ increases. The compound chondrule formation can occur only if the condition of $v_{\rm rel} < v_{\rm dest}$ is satisfied, where $v_{\rm dest}$ is the critical destruction velocity, otherwise ejectors will be disrupted upon impact. This condition is rewritten as
\begin{equation}
z < z_{\rm dest} \equiv \frac{ 2 r_{\rm l} \rho_{\rm mat} v_{\rm dest}^2 }{ 3 p_{\rm fm} } \left[ \left( \frac{ r_{\rm l} }{ r_{\rm s} } \right)^{1/2} - 1 \right]^{-2} .
\label{eq:z_max}
\end{equation}
From Eqs. (\ref{eq:z_min}) and (\ref{eq:z_max}), the region in which compound chondrules can be formed is $z_{\rm cool} < z < z_{\rm dest}$. In other words, the necessary condition for compound chondrule formation is given by $z_{\rm cool} < z_{\rm dest}$, which is rewritten as\footnote{We substitute $r_{\rm e} = r_{\rm l}$ in Eq. (\ref{eq:z_min}). }
\begin{equation}
\frac{ r_{\rm s} }{ r_{\rm l} } > \left( \frac{ 4 r_{\rm l} \rho_{\rm mat} v_{\rm dest} }{ 3 p_{\rm fm} t_{\rm cool} } + 1 \right)^{-2} .
\label{eq:non-destructive_collision}
\end{equation}
Substituting $r_{\rm l} = 300 \, {\rm \mu m}$, $p_{\rm fm} = 3 \times 10^4 \, {\rm dyne \, cm^{-2}}$, and $t_{\rm cool} = 0.08 \, {\rm s}$ to Eq. (\ref{eq:non-destructive_collision}), we obtain the appropriate size ratio of the small chondrule to large one as $r_{\rm s} / r_{\rm l} > 0.99$ for $v_{\rm dest} = 10^2 \, {\rm cm \, s^{-1}}$, $r_{\rm s} / r_{\rm l} > 0.91$ for $v_{\rm dest} = 10^3 \, {\rm cm \, s^{-1}}$, $r_{\rm s} / r_{\rm l} > 0.44$ for $v_{\rm dest} = 10^4 \, {\rm cm \, s^{-1}}$, and $r_{\rm s} / r_{\rm l} > 0.28$ for $v_{\rm dest} = 10^5 \, {\rm cm \, s^{-1}}$, respectively (see Fig. \ref{fig:sizeratio}). Ciesla (2006) mentioned that as chondrules cool they can survive collisions with one another at velocities up to $10^4 \, {\rm cm \, s^{-1}}$ due to viscous dissipation in the melt. If so, compound chondrule with the size ratio of about $\ga 0.5$ can be formed. However, it does not account for the mean value of $r_{\rm s} / r_{\rm l}$ from observations ($\sim 0.3$, Wasson et al. 1995). In order to explain the observation, it is required that $v_{\rm dest} \sim 1 \, {\rm km \, s^{-1}}$ or more, however, it seems unrealistic.  

In order to overcome this difficulty, we can consider the same possibilities as discussed in \S \ref{sec:efficiency}. For example, in the region shadowed from the ambient gas flow, $v_{\rm rel}$ does not increase with $z$ because ejectors are not accelerated. If $v_{\rm rel}$ just after entering the shadowed region is less than $v_{\rm dest}$, we obtain $z_{\rm dest} \rightarrow \infty$ because $v_{\rm rel}$ does not increase further. In this case, the compound chondrule formation is possible for arbitrary $r_{\rm s} / r_{\rm l}$. It implies that the compound chondrules with smaller value of $r_{\rm s} / r_{\rm l}$ can be formed in the shadowed region. In addition, we discussed the possibility that the disruption from partially-molten parent. In this case, $t_{\rm cool}$ can be shorter than that we assumed in this subsection. The shorter $t_{\rm cool}$ allows the compound chondrule formation for the wider range of the size ratio $r_{\rm s} / r_{\rm l}$ (see Fig. \ref{fig:sizeratio}). These issues will be discussed in detail in the forthcoming paper.

\hspace{1cm} {\bf [Figure \ref{fig:sizeratio}]}

\section{Summary}
\label{sec:summary}
We proposed a new scenario for compound chondrule formation named as ``fragment-collision model," in the framework of the shock-wave heating model. We modeled the disruption of molten cm-sized parent dust particle exposed to a high-velocity gas flow in order to estimate the efficiency of mutual collisions between small fragments assuming that all of them have the same radius. We obtained collision probability $P_{\rm coll}$ for a wide range of parameters (parent radius $r_{\rm p} = 0.2 - 5 \, {\rm cm}$, ejector radius $r_{\rm e} = 100 - 1000 \, {\rm \mu m}$, ram pressure of the gas flow $p_{\rm fm} = 10^4 - 10^5 \, {\rm dyne \, cm^{-2}}$, and viscosity of molten parent $\mu = 10 - 10^3 \, {\rm g \, cm^{-1} \, s^{-1}}$). The estimated collision probability was $\sim 0.1 - 1$ or more, which is about one order of magnitude larger than the observed fraction of compound chondrules. If we assume that all collisions lead to compound chondrule formation, this result suggests that our model can account for the observed fraction of compound chondrules when about $10\%$ or less of all chondrules formed via the fragmentation events. Since the fraction of chondrules which have undergone the fragmentation event depends on a size distribution of precursor dust particles in chondrule-forming region, it would be difficult to make a conclusion about the fraction of compound chondrules. However, numerical results of dust coagulation equations in the solar nebula seem to match well with our estimation (Nomura et al. 2007). In addition, our model does not require the dust enhancement in the pre-shock region because compound chondrules are formed from a single large dust particle. This is advantageous to explain the scarcity of isotopic fractionation of sulfur (Tachibana \& Huss 2005, Miura \& Nakamoto 2006). 

We also compared our model with other observational data. Akaki \& Nakamura (2005) measured the oxygen isotopic compositions of constituents of compound chondrules and found that in a three-isotopic diagram, all compound chondrules except for enveloping-types fall in the typical range obtained for single chondrules from the same chondrites. These observations are consistent with our model because two constituent chondrules are expected to originate from the same dust reservoirs. Gooding \& Keil (1981) and Akaki \& Nakamura (2005) reported that compound chondrules with non-porphyritic pairs are more frequent than that with porphyritic pairs. It is thought that these observations can be also explained by our model because the collision probability depends on the viscosity of molten parent dust particle. The dependence on the viscosity is consistent with the experimental results that porphyritic and non-porphyritic textures have been formed from partially- and completely-molten dust particles, respectively (Lofgren \& Russell 1986). Finally, the size ratios of secondary to primary in each set of compound chondrule have the mean value about 0.3 (Wasson et al. 1995). This result might be explained by our model because two fragments with different sizes are accelerated by the gas flow with different accelerations. As a result, these two fragments obtain large relative velocity and it would enhance the collision probability. Therefore, the compound chondrule of different-sized pair tends to be formed more frequently than that of same-sized one. 

However, it should be noted that all collisions do not necessarily lead to compound chondrule formation. For example, undesirably fast collisions cause disruption of ejectors upon impact. Assuming that the upper limit of the collisional velocity for coalescence is $10^4 \, {\rm cm \, s^{-1}}$ (Ciesla 2006), the appropriate size ratio of secondary (small) chondrule to primary (large) one is $r_{\rm s} / r_{\rm l} \ga 0.5$, which does not account for the observations (mean value of $r_{\rm s} / r_{\rm l} \sim 0.3$, Wasson et al. 1995). In addition, ejectors should cool moderately before collision to make compound chondrule not to fuse into a single droplet. The collision probability after the moderate cooling, however, is much smaller than that of total collisions. In order to overcome these difficulties, we consider other physics that we did not take into account in this paper (e.g., the gas flow is blocked by the parent and/or numerous numbers of ejectors). These issues will be discussed in the forthcoming paper. In addition, our model should be tested by other methods, e.g., three-dimensional (thermo-)hydrodynamics simulation or aerodynamic disruption experiment in the future.

\section*{Acknowledgment}
We are grateful to Drs. Tomoki Nakamura and Fred J. Ciesla, and an anonymous referee for useful comments in this study. H.M. and S.Y. were supported by the Research Fellowship of Japan Society for the Promotion of Science for Young Scientists. T.N. was partially supported by the Ministry of Education, Science, Sports, and Culture, Grant-in-Aid for Scientific Research (C), 1754021.

\appendix

\section{Hydrodynamics in Liquid Layer}
\label{sec:kato2006}
\begin{figure}[p]
\center
\includegraphics[scale=1.2, angle=0.]{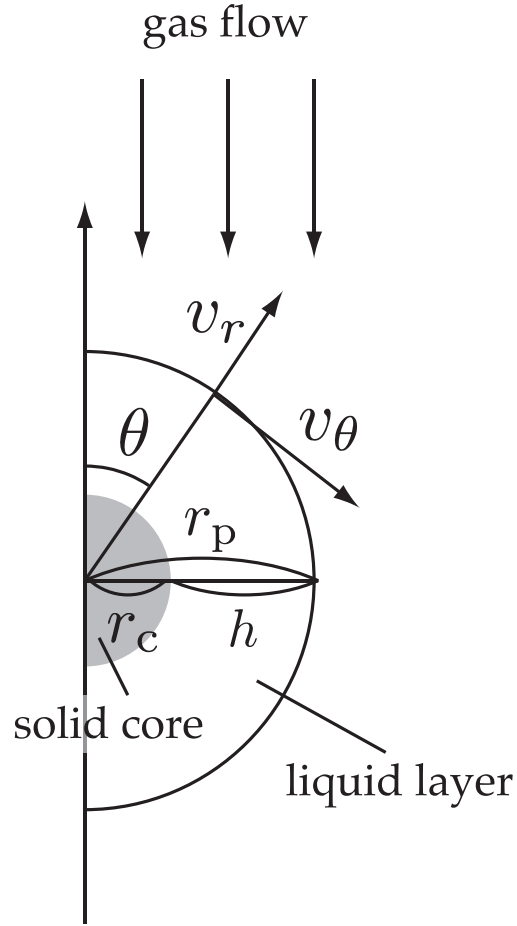}
\caption{Schematic picture of the set-up in Kato et al. (2006) to solve the hydrodynamics equations of molten dust particle. A solid core (gray) of radius $r_{\rm c}$ is surrounded by a liquid layer (white) of width $h$, where the dust radius $r_{\rm p} = r_{\rm c} + h$. The radial and tangential fluid velocities inside the liquid layer, $v_r$ and $v_{\theta}$, were obtained as a function of $r$ and $\theta$, where $r$ is a distance from the center. The ambient gas flows from up to bottom in this figure.}
\label{fig:kato_model}
\end{figure}

\begin{figure}[p]
\center
\includegraphics[scale=.7, angle=-90.]{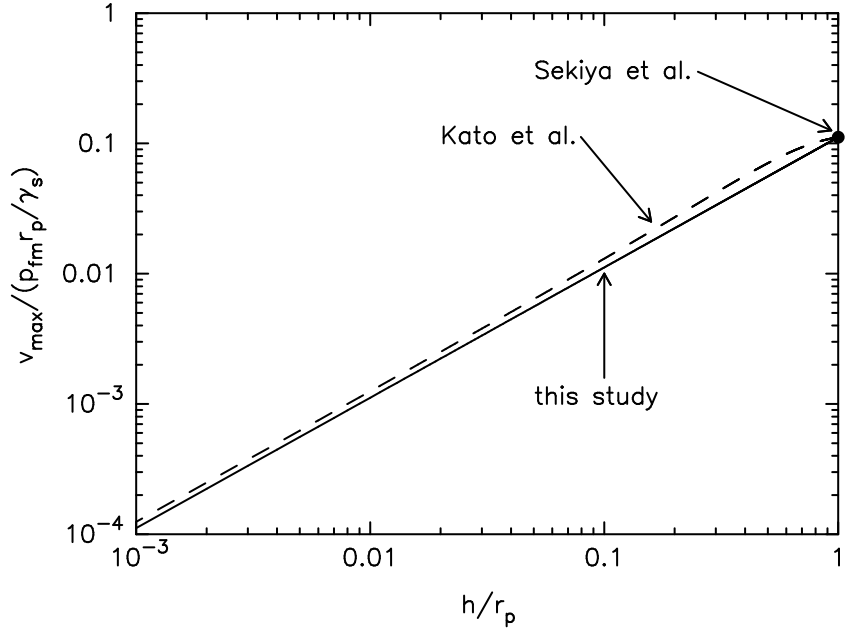}
\caption{Maximum tangential velocity at the surface of liquid layer $v_{\rm max}$ as a function of a width of liquid layer $h$. The solution obtained by Sekiya et al. (2003) was for a completely-molten particle, so it corresponds to $h / r_{\rm p} = 1$ as pointed by a filled circle. The solutions for Kato et al. (2006, dashed curve) were for arbitrary values of $h / r_{\rm p}$ from 0 to 1 and $v_{\rm max} \rightarrow 0$ for $h / r_{\rm p} \rightarrow 0$. We linearly interpolate from $v_{\rm max} = 0$ at $h / r_{\rm p} = 0$ to $v_{\rm max} = 0.112 \, p_{\rm fm} r_{\rm p} / \mu$ at $h / r_{\rm p} = 1$ (solid line). }
\label{fig:kato}
\end{figure}

Kato et al. (2006) examined the hydrodynamics of the liquid layer by analytically solving the hydrodynamics equations for a core-mantle structure with a linear approximation. Fig. \ref{fig:kato_model} shows a schematic picture of the set-up in their analysis. According to their solutions, we can obtain the radial and tangential velocities, $v_{r}$ and $v_{\theta}$, at arbitrary position in the liquid layer. The tangential velocity has the maximum value $v_{\rm max}$ at $\theta = 0.265 \pi$ on the surface of liquid layer. Fig. \ref{fig:kato} shows $v_{\rm max}$ as a function of the layer width $h$. The horizontal axis is a normalized layer width $h / r_{\rm p}$ and the vertical one is a normalized velocity $v_{\rm max} / (p_{\rm fm} r_{\rm p} / \mu)$. The solution obtained by Kato et al. (2006) is a complex function (dashed), however, it can be approximated by a linear interpolation between $v_{\rm max} = 0$ for $h / r_{\rm p} = 0$ and $v_{\rm max} = 0.112 \, p_{\rm fm} r_{\rm p} / \mu$ for $h / r_{\rm p} = 1$. The value of $v_{\rm max}$ for $h / r_{\rm p} = 1$ corresponds to the solution obtained by Sekiya et al. (2003), in which they analyzed the hydrodynamics of a completely-molten dust particle. In our model, we adopt the linear interpolation for $v_{\rm max}$ (solid line) as the velocity dispersion of ejectors disrupted from a molten parent particle, which is given by
\begin{equation}
v_{\rm max} = 0.112 \, p_{\rm fm} h / \mu .
\end{equation}

\hspace{1cm} {\bf [Figure \ref{fig:kato_model}]}

\hspace{1cm} {\bf [Figure \ref{fig:kato}]}

\clearpage

\clearpage

\clearpage

\clearpage

\clearpage

\clearpage

\clearpage

\clearpage

\end{document}